\documentclass[runningheads]{llncs}

\usepackage{float}
\titlerunning{autopet II Solution}
\usepackage{caption}
\usepackage[T1]{fontenc}
\usepackage{orcidlink}
\usepackage{amsfonts}
\usepackage{graphicx}
\begin{document}

\title{AutoPETIII: The Tracer Frontier \newline \textit{What frontier ?}}
%
%

\author{
Zacharia Mesbah\inst{1,2,4} \and
Leo Mottay\inst{1,2,4} \and
Romain Modzelewski\inst{2}\and
Pierre Decazes\inst{2}\and
Sébastien Hapdey\inst{2}\and
Su Ruan\inst{1}\and
Sebastien Thureau\inst{3} \\
}

\authorrunning{}
%
\institute{INSA Rouen Normandie, Univ Rouen Normandie, Université Le Havre Normandie, Normandie Univ, LITIS UR 4108, F-76000 Rouen, France\\
 \and 
Nuclear Medicine Department, Henri Becquerel Cancer Center, Rouen, France\\ 
\and 
Radiotherapy Department, Henri Becquerel Cancer Center, Rouen, France \\ \and
Siemens Healthineers\\
\email{zacharia.mesbah@chb.unicancer.fr}
}

\maketitle   
\begin{abstract}

For the last three years, the AutoPET competition gathered the medical imaging community around a hot topic: lesion segmentation on Positron Emitting Tomography (PET) scans. Each year a different aspect of the problem is presented; in 2024 the multiplicity of existing and used tracers was at the core of the challenge. Specifically, this year's edition aims to develop a fully automatic algorithm capable of performing lesion segmentation on a PET/CT scan, without knowing the tracer, which can either be a FDG or PSMA-based tracer.
In this paper we describe how we used the nnUNetv2\cite{isensee2018nnunet} framework to train two sets of 6 fold ensembles of models to perform fully automatic PET/CT lesion segmentation as well as a MIP-CNN to choose which set of models to use for segmentation.

\end{abstract}

\section{Introduction}
For cancer detection and diagnosis, Positron Emission Tomography (PET) imaging is extremely valuable. It allows to explore specific functions in the body, which is especially fit for cancer detection since cancer cells are cells exhibiting abnormal behavior. PET consists in injecting a radioactive tracer in the patient's body and observing its distribution using the gamma rays it emits.

Tracers are designed to target a specific function of the body. The most common, ${}^{18}F$-FluoroDesoxyGlucose (${}^{18}$FDG) is a sugar, which accumulates (i.e. uptakes) in the parts of the body that consume energy (i.e. high-function). The cancer cells, in their uncontrolled reproductive frenzy, need a lot of energy. They're usually very visible on PET scans, which makes this imaging modality the keystone of cancer detection and cancer treatment protocols.

It is always accompanied by a Computed Tomography (CT) scan, which serves for attenuation correction of the PET scan. However, the images from this modality also provide a different, useful information. They show the mapping of density in the body, which gives anatomical information. The combination of the anatomical and functional information is used by physicians to determine which uptakes are malignant (lesions) and which are not (physiological uptake). Some examples for non-malignant uptakes of FDG are liver, brain and kidneys. The liver and kidneys have high uptake because of their cleaning role, while the brain is always active and thus consumes sugar.


Over the last decade, another family of tracers has been used more and more. The Prostate Specific Membrane Antigen (PSMA) is a protein usually found in both healthy and cancerous prostate tissues. Due to this property, PSMA has been used in nuclear medicine to track prostate cancer in the human body. Unlike FDG, this family of tracer does not accumulate in the organs with the most activity: a healthy brain will show no activity on a PSMA PET scan. However, the waste clearing organs will still show activity as well as other organs like lacrimal glands and the gastro-intestinal tract.

FDG and PSMA PET scans are slowly anchoring themselves in routine clinical practice, which means an increase in the work intensive  task of PET analysis by physicians. The first step of this task is to detect and delineate all the malignant lesions. Indeed the number of lesions and their localization are key in establishing patient prognosis and determining further treatment.

To alleviate this task from the physicians' workload, to develop a reliable automatic PET lesions segmentation tool would be extremely helpful.

\section{Competition}

\subsubsection{This year's edition,}

AutoPET III challenge aims, just as its predecessors did, to build an algorithm capable of delineating cancerous lesions in PET/XCT scans. However this year's novelty is the appearance of not only FDG but also  PSMA PET scans. The catch is that no prior knowledge is given (at test time) about the tracer.

To achieve this, participants are given 1014 FDG PET/CT scans and 597 PSMA PET/CT scans, gathered from two German medical centers:
\begin{itemize}
    \item University Hospital of Tübingen (FDG data)
    \item University Hospital of the LMU in Munich (PSMA data)
\end{itemize}

For each of these studies, two modalities are available, in NifTi format:
\begin{itemize}
    \item The Computed Tomography, resampled towards the PET spacing
    \item The PET, for which voxel values have been converted to Standardized Uptake Value.
\end{itemize}   

Studies selected were from patients with lung cancer, lymphoma, melanoma or healthy patients for FDG and prostate cancer patients for PSMA.

\newpage

\subsubsection{Dataset}
The dataset is split in three parts:

\begin{itemize}
    \item The training set, which can be used by all contestants to build and train their algorithm.
    It corresponds to the 1611 studies presented above.
    \item The preliminary test set, containing 5 scans. Considering the statistical non-significance of this set, it should be used to make sure no catastrophic failure happens in the algorithm execution.
    \item The final testing set, with 200 scans. This is the scoring set which will be used for the competition ranking.
\end{itemize}

\begin{table}
\centering
\begin{tabular}{|c|c|c|c|}
\hline
& \textbf{Training set} & \textbf{Valid}   & \textbf{Test} \\
\hline
\textbf{No. of Scans} & 1611 & 5 & 200\\
\hline
\end{tabular}
\caption{Dataset Distribution: train / test split}
\label{table:distribution}
\end{table}

\subsubsection{Evaluation}

Evaluation on the final test set is performed using a mixed model. The metrics included are:
\begin{itemize}
    \item Dice Similarity Coefficient (DSC), a geometric similarity metric
    \item False Positive Volume: The number of pixels in isolated, connected components of the prediction which are false positives
    \item False Negative Volume: The number of pixels in isolated, connected components of the ground truth which are false negatives 
\end{itemize}

\subsubsection{nn-Unetv2}
nnUNetv2 is a deep neural network training framework, allowing anyone to train out-of-the-box segmentation networks, in two and three dimensions. It determines automatically the model architecture parameters (depth, width) and the training hyper-parameters using heuristics.
nnUNetv2 is displaying results at the state-of-the-art to this day as described in a recent paper\cite{isensee2024nnunetrevisitedrigorousvalidation}.

It is considered an essential baseline for any research conducted on semantic segmentation, in the medical imaging domain mainly. It is also used in most winning submissions to MICCAI and other challenges based around segmentation.

\newpage

\section{Our Method}
\subsubsection{Windowing}
We used a CT and PET windowing method. Instead of using a 2 channels input  to our network, we use 4 channels. The first two channels remain the CT and PET volumes, but the third and fourth channels are clipped versions of the CT and PET scans. Precisely:
\begin{itemize}
    \item The PET scan is clipped between 0 and 20 SUV
    \item The CT scan is clipped between -300 and 400 HU
\end{itemize}

This windowing's goal is to deal with two problems caused by input data normalization:
\begin{itemize}
    \item By normalizing between min and max, the values are squashed, which means that slow variations in the image are represented with very similar values which can in turn make differentiation harder for the network. We normalize over a smaller range of values to limit this effect
    \item For each patient the maximum value can be different (especially in PETs). This lowers the meaningfulness of the normalized voxel values inputted in the network, as it is also dependant on the volume's maximum value. Windowing also reduces this effect by making sure we are in a specific range of values
\end{itemize}

\subsubsection{Supplementary labels}
We re-used a method from last year's AutoPET II competition. The winning team had drastically reduced the amount of false positives by adding other anatomical contours to the segmentation. We defined a list of organs that were interesting for PET segmentation:
\begin{itemize}
    \item the brain, heart and aorta : high FDG uptake
    \item liver, kidneys, urinary bladder: waste clearing activity
    \item spleen: sometimes very active
    \item  digestive system, prostate: high PSMA uptake
    \item skeleton,
    \item lungs, pancreas: added localization for the remaining organs
\end{itemize}

\begin{figure}
    \centering
    \includegraphics[width=0.265\linewidth]{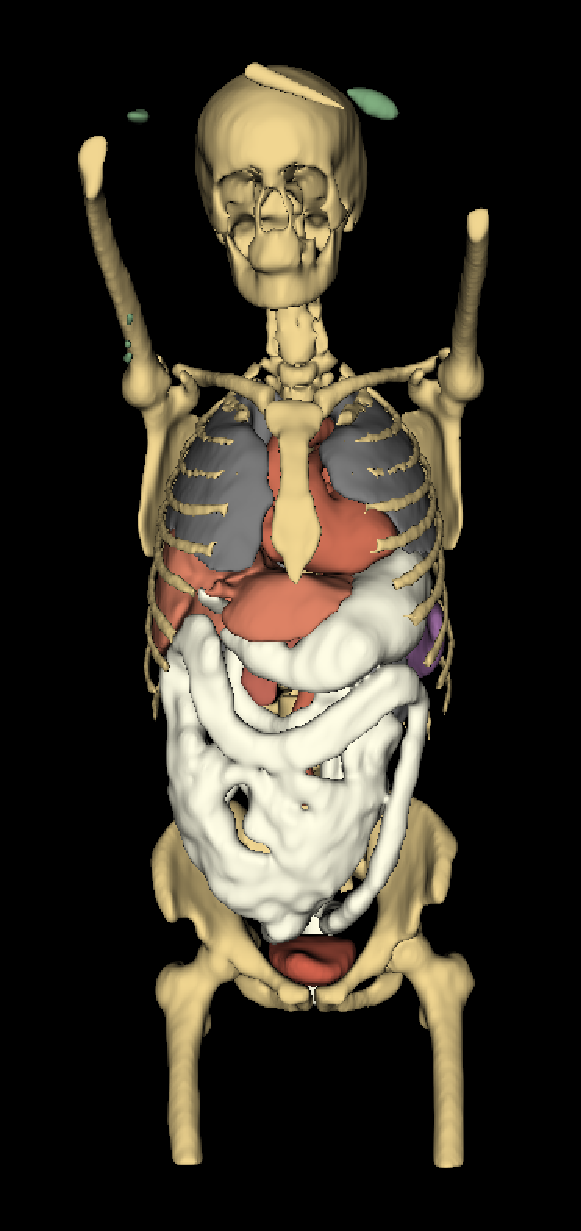}
    \includegraphics[width=0.19\linewidth]{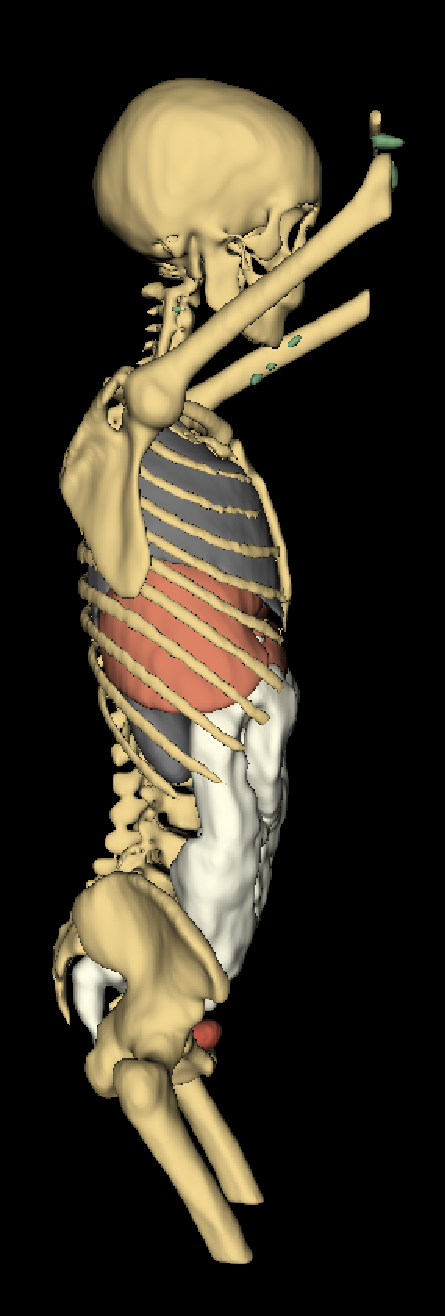}
    \includegraphics[width=0.25\linewidth]{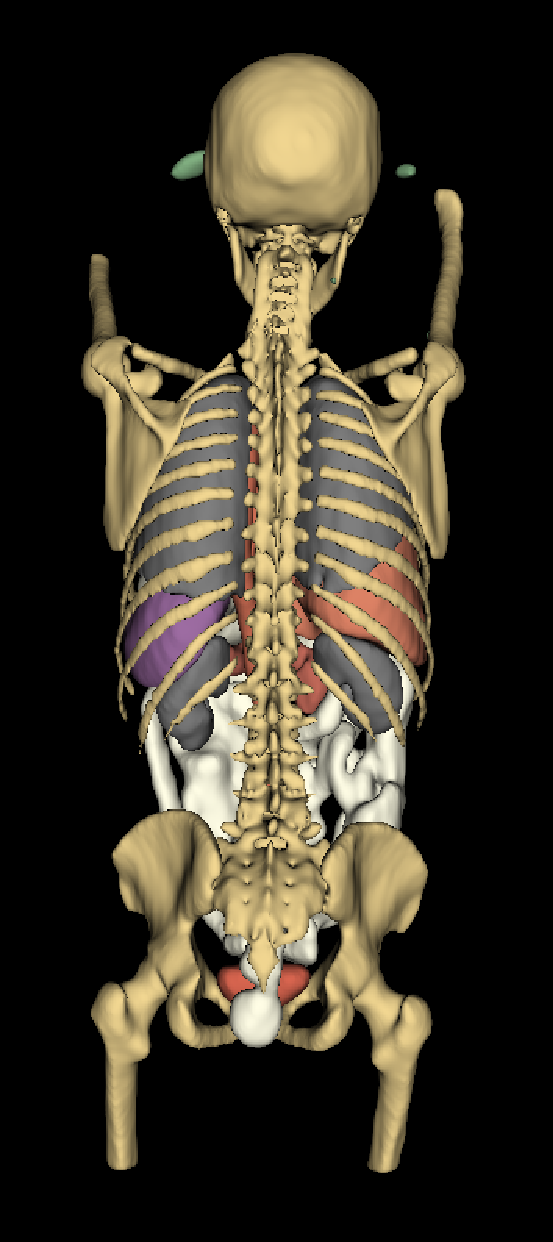}
    \includegraphics[width=0.2185\linewidth]{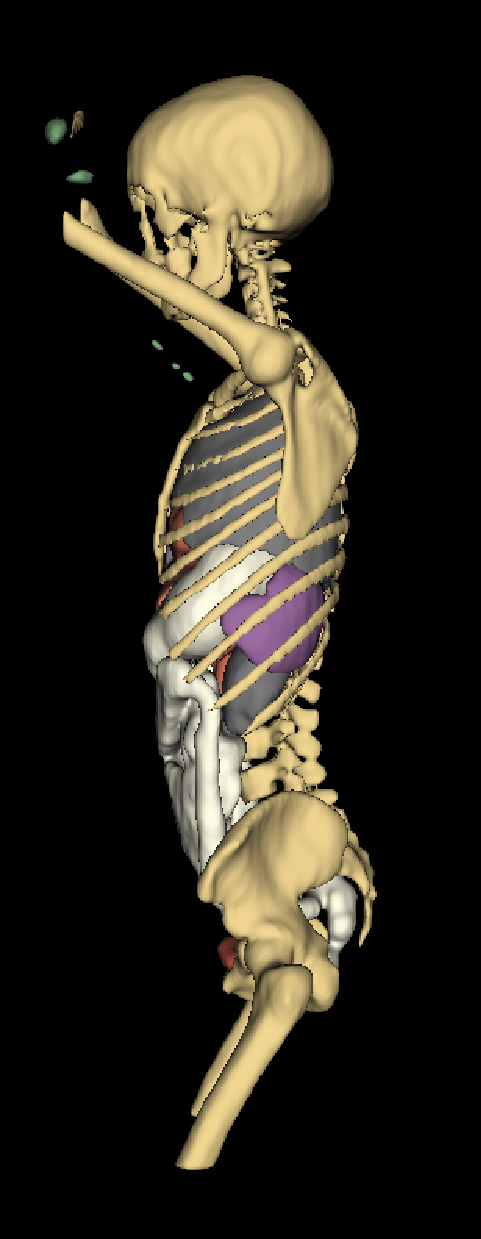}
    \caption{Example of added labels for a FDG patient}
    \label{fig:supplementary_labels}
\end{figure}

We used TotalSegmentator\cite{Wasserthal_2023} to perform segmentation for all of these organs using the CT scans. We had to group some organs together to form broader classes (bones, digestive system, lungs etc...) in order to reduce the computational overhead.
On top of these organs we added the lesion back with maximum priority.

This method's goal is to help the network figuring out malignant uptake from benign uptake.

\newpage

\subsubsection{Tracer Discriminator}

Our approach consists in using tracer-specific neural networks for tumor delineation. However, since the injected radio-tracer is not disclosed alongside volumes, a tracer detection strategy is required. To address this, we use a neural network for tracer discrimination. Its architecture consists of six consecutive 2D convolutions layers, followed by five fully connected layers, all with ReLU activation and a final sigmoid activation.
Initially, the volumes are resampled (3x3x3 spacing) and their maximum intensity projection (MIP) is computed in the coronal plane.
Next, a zero-padded 224x224 window is extracted around the center of the MIP and fed to our neural network.
During training, the model’s outputs (predicted tracer types) are compared against ground truth using Binary Cross-Entropy (BCE) as the loss function.
Optimization is performed using AdamW, with an initial learning rate of $1e^{-4}$, over 100 epochs with early stopping to reduce over-fitting.
Inference times on the entire dataset, tested on an Intel i7-11850H CPU, averages to 2.18 seconds per patient and 5-folds cross-validation yielded 99.64\% accuracy.

\newpage

\subsubsection{Segmentation Networks}

As described earlier, we use two tracer-specific segmentation networks. Both are trained using the nnUNet framework, with mostly unchanged parameters/hyper-parameters.
The only change occurred in the setting for the FDG model. Indeed, nnUNet is using the median spacing as a target spacing for all training data. However, the median (3mmx2mmx2mm) spacing for FDG data implied too much inference time as this was limited to 5 minutes per patient. We chose to replace it by an isotropic larger spacing (3.3mmx3.3mmx3.3mm), which allowed us to run  a 6 folds ensembling with Test-Time Augmentation (TTA) for most patients \footnote{We still set a threshold on the number of voxels where we remove some TTA to stay in the time limit.}.

We then trained the regular 5 folds ensemble with no post-processing for both tracers.

\section{Results}

\begin{table}[H]
\begin{center}
    
\begin{tabular}{|c|c|c|c|}
\hline
\textbf{}& \textbf{Dice Score(\%)} & \textbf{FNV} & \textbf{FPV}  \\ \hline
nnUNet baseline (1 model) &  61.41 & 25.61 & 18.71 \\ \hline
\textbf{Ours} &  74.91 & 40.72 & 0.760 \\ \hline

\end{tabular}
\caption{Results of our model on the preliminary test set}
\label{table:results}
\end{center}
\vspace{-10mm}
\end{table}
\noindent Our model obtained results that situated us middle of the pack on the preliminary test set, which does give us an idea of what our final result will be.

\section{Conclusion}
The key takeaway from our participation in this challenge is that a dataset of such dimensions makes the training of highly reliable PET lesions segmentation deep learning based tools accessible to researchers in cancer research centers.
We will conduct future work, as well as review the others participants' methods to attempt developing a robust PET segmentation model.

Such a model would alleviate the burden of PET segmentation off the shoulders of physicians, in turn freeing their time to conduct more impactful work.

%
%
%
\bibliographystyle{unsrt}
\bibliography{mybibliography}

\begin{thebibliography}{1}

\bibitem{isensee2018nnunet}
Fabian Isensee, Jens Petersen, Andre Klein, David Zimmerer, Paul~F. Jaeger, Simon Kohl, Jakob Wasserthal, Gregor Koehler, Tobias Norajitra, Sebastian Wirkert, and Klaus~H. Maier-Hein.
\newblock nnu-net: Self-adapting framework for u-net-based medical image segmentation, 2018.

\bibitem{isensee2024nnunetrevisitedrigorousvalidation}
Fabian Isensee, Tassilo Wald, Constantin Ulrich, Michaal Baumgartner, Saikat Roy, Klaus Maier-Hein, and Paul~F. Jaeger.
\newblock nnu-net revisited: A call for rigorous validation in 3d medical image segmentation, 2024.

\bibitem{Wasserthal_2023}
Jakob Wasserthal, Hanns-Christian Breit, Manfred~T. Meyer, Maurice Pradella, Daniel Hinck, Alexander~W. Sauter, Tobias Heye, Daniel~T. Boll, Joshy Cyriac, Shan Yang, Michael Bach, and Martin Segeroth.
\newblock Totalsegmentator: Robust segmentation of 104 anatomic structures in ct images.
\newblock {\em Radiology: Artificial Intelligence}, 5(5), September 2023.

\end{thebibliography}

\end{document}